\documentclass[twocolumn]{jpsj2}

\def\runtitle{A Consistent Picture of a Collapsing Bose-Einstein Condensate}
\def\runauthor{MASAHITO UEDA AND HIROKI SAITO}

\title{A Consistent Picture of a Collapsing Bose-Einstein Condensate}

\author{Masahito Ueda and Hiroki Saito}

\inst{%
Department of Physics, Tokyo Institute of Technology,
Meguro-ku, Tokyo 152-8551, Japan, and\\
CREST, Japan Science and Technology Corporation (JST), 
Saitama 332-0012, Japan}

\recdate{\today}

\abst{
We present a consistent picture of the recent JILA experiments on the collapse of a Bose-Einstein condensate (BEC) [E. A. Donley, et al., Nature {\bf 412}, 295 (2001)] based on a generalized Gross-Pitaevskii equation that incorporates the effect of a three-body loss.
The questions we address in this paper are: (1) Why does a BEC survive, albeit partially, after the collapse? (2) Why does the number of the remnant BEC atoms observed at JILA exceed the critical value?  (3) What is the origin of the atom burst? (4) What is the origin of the atom jet and is it coherent?
We review our predictions on pattern formation in the course of collapse and on the split instability of vortices that can test our collapsing theory.
}

\kword{
Bose-Einstein condensation, collapsing dynamics, Feshbach resonance, vortices
}

\begin{document}
\sloppy
\maketitle

\section{Why attractive interaction?}

One of the unique features of gaseous Bose-Einstein condensates (BECs) is their unprecedented controllability when applied to the creation, manipulation, or probing of the BEC system, and as a result, many previously gedanken experiments have become or are expected to become a reality. A prime example is the dynamics of a collapsing BEC, which mimic some of those of collapsing and exploding stars~\cite{Nature}. 

In a gaseous BEC system, the interaction between atoms is characterized by the s-wave scattering length $a$.
At zero magnetic field, 
$^7$Li~\cite{Bradley}, $^{85}$Rb~\cite{Cornish}, and $^{133}$Cs~\cite{Weber} have negative scattering lengths, which implies that atoms interact attractively and the condensates tend to collapse upon themselves.
We may thus expect the density instability and the ensuing collapsing dynamics of the condensate which are similar to those of supernova explosions.
The attractive interaction also drastically changes the nature of superfluidity
and that of topological excitations.

In the present paper we present a consistent picture~\cite{SaitoPRL01,SaitoPRA01,SaitoPRA02} of the so-called Bosenova experiments recently performed at JILA~\cite{Donley}. Our theory is based on a generalized Gross-Pitaevskii (GP) equation that incorporates the effect of a three-body recombination loss~\cite{Kagan98}.Theoretical analyses on the Bosenova experiments using similar approaches~\cite{Santos,Adhikari,Savage} --as well as analyses based on elastic condensate collisions~\cite{Duine} 
and molecular formation~\cite{Milstein,Mackie,Kohler}--
have also been reported.
  We shall review some new phenomena on pattern formation in the course of collapse~\cite{SaitoPRL01} and on the split instability of vortices~\cite{SaitoPRL02} that our theory predicts and hence can test our collapsing theory.

\section{Controlled collapse of the condensate}

It has been well established that an attractive BEC may exist
in a confined system when the number of BEC atoms is below a
certain critical value~\cite{Bradley,Ruprecht,Roberts}. 
In this case, the zero-point energy due to confinement serves as a 
kinetic barrier against collapse, allowing formation of a metastable BEC.
It is through this dynamically generated barrier that the system 
is believed to tunnel when the number of BEC atoms is just below 
the critical value and the system collapses spontaneously~\cite{Ueda98,Stoof}.

The subject of the present paper is the controlled collapse 
of the condensate~\cite{Cornish,Roberts,Donley,Gerton}.
Recently, the JILA group utilized the Feshbach resonance to switch
the sign of interactions from repulsive to attractive, and caused
a BEC to collapse~\cite{Cornish,Roberts,Donley}.
After the switch, the atomic cloud shrank and eventually disappeared 
because the cloud became too small. 
That is, the BEC collapsed. 
Surprisingly, however, just after the collapse they observed a burst 
of atoms emanating from the remnant BEC~\cite{Donley}, a phenomenon similar 
to supernova explosion. 
Before this experiment was performed, there had been controversy as 
to what happens to a BEC once it begins to collapse. 
Some researchers predicted a complete collapse, 
while others~\cite{Kagan98,SaitoPRL01} predicted a partial one.
The JILA experiments clearly demonstrated that the collapse is partial.

To understand the underlying physics of this partial collapse, 
let us consider a BEC confined in an isotropic harmonic trap. 
The dynamics of a BEC in a dilute-gas regime are well described by the 
GP equation
\begin{eqnarray}
i\frac{\partial}{\partial t}\Psi=-\frac{1}{2}\nabla^2\psi
+\frac{1}{2}(x^2+y^2+\lambda z^2)\psi
+g|\psi|^2\psi,
\label{GP}
\end{eqnarray}
where $\psi$ represents the order parameter 
or the ^^ ^^ wave function" of the condensate, normalized
as $\int|\psi|^2d{\bf r}=1$, and $\lambda$ is the asymmetry paramter 
of the trap.
In Eq.~(\ref{GP}), the time and the length are measured in units of the inverse trap frequency $\omega^{-1}$ and the natural length scale of the harmonic trap 
$d_0\equiv(\hbar/m\omega)^{1/2}$, respectively, where $m$ is the atomic mass.
The last term in Eq.~(\ref{GP}) describes the mean-field interaction
between atoms whose strength $g$ is given in terms of the number of BEC atoms
$N_0$ and the s-wave scattering length $a$ as $g=4\pi N_0 a/d_0$.

\begin{figure}[t]
\begin{center}
\includegraphics[width=8cm]{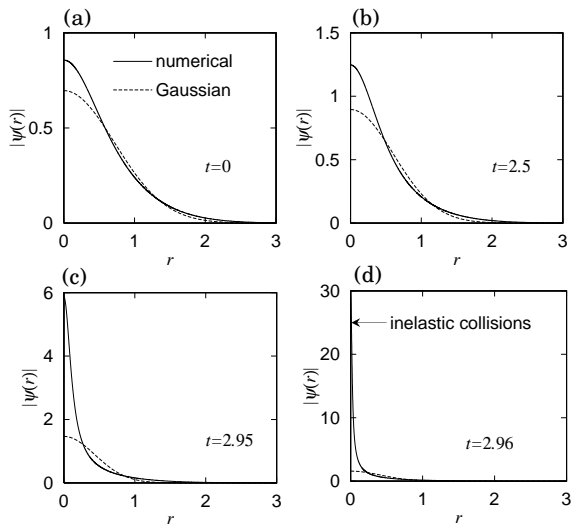}
\end{center}
\caption{The solid curves show the profiles of the wave functions
$|\psi(r,t)|$ at $t=0, 2.5, 2.95$, and 2.96, obtained by numerically
integrating the Gross-Pitaevskii equation~(\ref{GP}) with $\lambda=1$. 
The dashed curves indicate Gaussian fits that have the same widths as the exact solutions.
}
\label{f1}
\end{figure}

Figure~\ref{f1} shows the time development of the collapsing condensate wave function~\cite{SaitoPRA01}obtained by numerically integrating Eq.~(\ref{GP})
using a finite-difference method with the Crank-Nicholson scheme~\cite{Ruprecht}.
The superimposed dashed curves represent Gaussian fits having the same width 
as the numerical ones.
At $t = 0$, we increase the strength of attractive interaction to
just above its critical value and cause the system to collapse 
[Fig.~\ref{f1}(a)].   
At $t=2.5$, the peak height grows significantly, but 
the wave function fairly well maintains the Gaussian form
[Fig.~\ref{f1}(b)].
However, at $t=2.95$, the peak height grows very rapidly and the Gaussian 
approximation clearly breaks down [Fig.~\ref{f1}(c)].
Remarkably, just a little after this ($t=2.96$), 
we see that the peak height has grown sufficiently that inelastic 
collisions would be substantial, while 
in most other regions inelastic collisions are negligible
[Fig.~\ref{f1}(d)].
This inelastic-collision-dominated regime is extremely localized
and occupies only one millionth of the condensate volume.
Thus the condensate has an extremely small atom 
drain --or a tiny black hole-- at its center. 

\medskip
{\it 
That the collapse occurs only in an extremely localized region and nowhere else in the BEC is the primary reason that the collapse occurs only partially and we have a remnant BEC.}

\section{What value should we take for $K_3$?}

To follow the dynamics of the system after the collapse, 
we include in the GP equation the atomic loss due to the 
three-body recombination loss~\cite{Kagan98}:
\begin{eqnarray}
i\frac{\partial}{\partial t}\Psi=-\frac{1}{2}\nabla^2\psi
+\frac{1}{2}(x^2+y^2+\lambda z^2)\psi
-\frac{i}{2}K_3|\psi|^4\psi,
\label{GP_loss}
\end{eqnarray}
where $K_3$ is the three-body recombination loss rate coefficient.
(The two-body dipolar loss has little effect on the collapsing dynamics.)

The collapsing dynamics depend crucially on the value of $K_3$ 
--the three-body recombination rate coefficient. 
However, no direct measurement of $K_3$ has been reported 
near the Feshbach resonance peak.
It is therefore important to devise a method for determining $K_3$ 
from other experimentally measurable quantities. 
For this purpose, we note that the peak density in the 
course of collapse ceases to grow when the interaction term
is counterbalanced by the three-body loss term, i.e., 
$g|\psi|^2\sim K_3|\psi|^4/2$, and hence we have
$|\psi|^2\sim 2g/K_3$.
As explained in the next section, the atom burst is caused by
a slight imbalance between the kinetic energy and the attractive
interaction. 
We may thus expect that the mean burst energy is proportional to 
the interaction energy at the time when those two balance, i.e., 
$g|\psi|^2\sim 2g^2/K_3$.

\begin{figure}[t]
\begin{center}
\includegraphics[width=8cm]{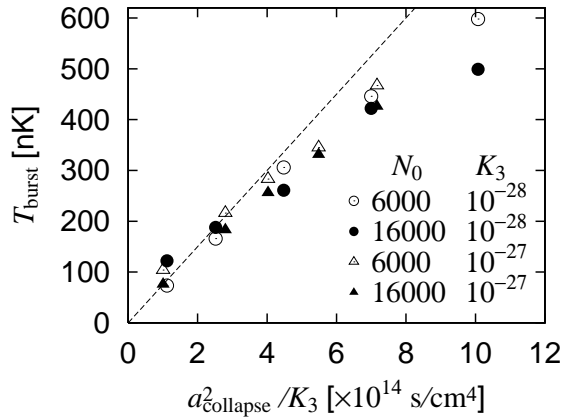}
\end{center}
\caption{
The mean burst energies $k_{\rm B}T_{\rm burst}$ 
plotted against $a_{\rm collapse}^2/K_3$ for various
values of $a_{\rm collapse}$ and $K_3$, where $k_{\rm B}$ is the Boltzmann
constant and $a_{\rm collapse}$ is the value of the s-wave scattering 
length $a$ when the collapse occurs.
The dashed line is a theoretical fit (\ref{mbe}) 
obtained in Ref.~\cite{SaitoPRA01}
}
\label{f2}
\end{figure}

To check this hypothesis, we have performed extensive numerical    
simulations by varying $g$ and $K_3$ and calculating the mean burst
energy for each set of $g$ and $K_3$.
We calculate the energy distribution of the burst at     
$t=t_{\rm implosion}+\pi/2$ when the burst atoms spread  
maximally and hence can easily be distinguished from the 
remnant BEC.
Figure~\ref{f2} summarizes our numerical simulations.

From Fig.~\ref{f2} we find that when $T_{\rm burst}<200$nK the 
mean burst energy is indeed proportional to 
$g^2/K_3\propto a^2_{\rm collapse}/K_3$ and is given by 
\begin{eqnarray}
{\rm mean \ burst \ energy}\simeq 
2.6\hbar^3\frac{a_{\rm collapse}^2}{m^2K_3}.
\label{mbe}
\end{eqnarray}
By taking the experimentally measured mean burst energy $\sim$100nK, 
we obtain  
$K_3\simeq 2\times 10^{-28}$ cm$^6$/s,
which we will use in the following discussions. 

\medskip
{\it All relevant parameters have thus been determined experimentally
and our theory includes no adjustable parameters.}

\section{Intermittent implosions}

\begin{figure}[t]
\begin{center}
\includegraphics[width=8cm]{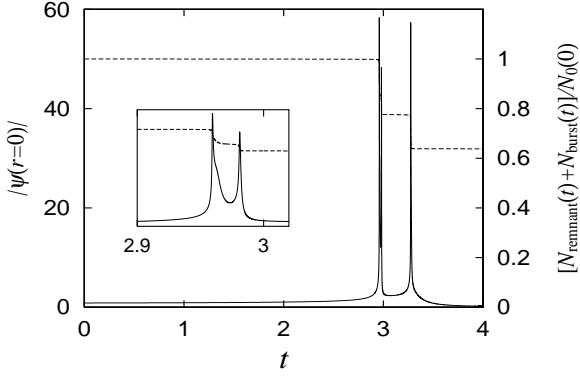}
\end{center}
\caption{
Time evolution of the peak height $|\psi(r=0,t)|$ of the collapsing
wave function according to Eq.~(\ref{GP_loss}) with $\lambda=1$.
The dashed curves (referring to the right axis) show the fraction of 
the atoms remaining in the trap, which is defined as the ratio of the
number of remnant BEC atoms $N_{\rm remnant}(t)$ plus the number of burst
atoms $N_{\rm burst}(t)$ at time $t$ against the initial number of BEC atoms
$N_0(0)$.
The inset shows an enlarged view of the intermittent implosions.
}
\label{f3}
\end{figure}

The solid curve in Fig.~\ref{f3} shows the time evolution of the
peak height of the wave function~\cite{SaitoPRL01}.
In the initial stage of the collapse, the atomic density grows   
very slowly. However, a rapid implosion breaks out at $t\simeq2.96$;
a blowup of this implosion is shown in the inset.
It is interesting to observe that implosions occur not once but 
several times intermittently, and at each implosion only several
tens of atoms are lost from the condensate. 
This is a consequence of the fact that
the collapse occurs in an extremely localized region.

\medskip
The physics of the intermittent implosions may be understood as follows.

\medskip
\begin{itemize}
\item 
Initially, the attractive interaction is made to dominate the kinetic 
energy, so the condensate shrinks towards the central region and the  
peak height grows until the growth is stopped by inelastic collisions.
Only the localized high density region collapsed and disappeared due  
to the atomic loss.

\item 
After the atoms at the density peak are suddenly removed due to the 
three-body loss, the remaining BEC still peaks steeply and 
the zero-point kinetic pressure arising from Heisenberg's uncertainty
principle barely surpasses the attractive interaction.
Thus the central portion of the condensate subsequently expands due to this
extra zero-point pressure.
{\it This slight surplus of kinetic energy is the origin of the atom 
burst and explains why the energy of the burst atoms is as low as 100nK.}

\item As soon as the atom burst occurs, the peak height decreases
very rapidly, which makes the attractive interaction again dominate
the zero-point pressure, thereby inducing the subsequent implosions.

\item This implosion-burst cycle repeats, giving rise to {\it intermittent
implosions}, until the number of the 
remaining atoms becomes so small that the attractive force no longer 
overcomes the kinetic pressure.
\end{itemize}

\begin{figure}[t]
\begin{center}
\includegraphics[width=8cm]{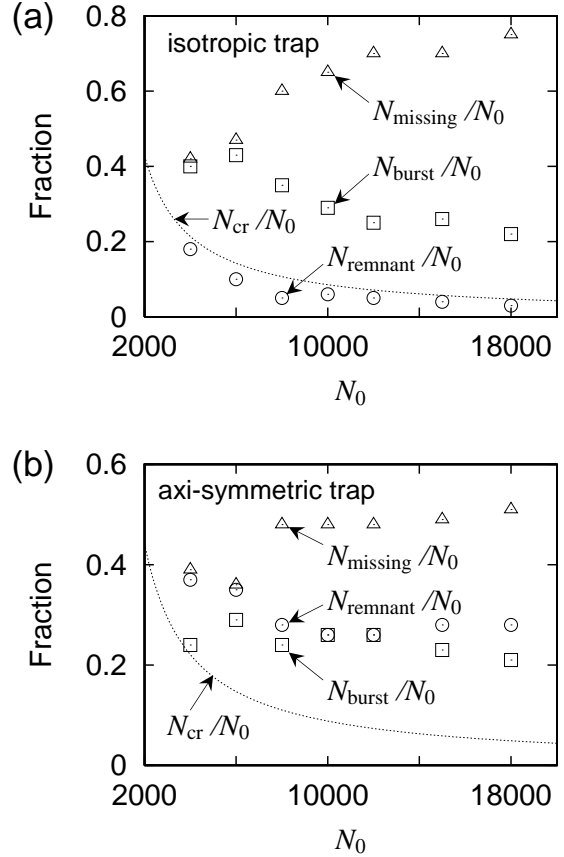}
\end{center}
\caption{
Dependences on the initial number of atoms $N_0$ of
the remnant $N_{\rm remnant}$, burst $N_{\rm burst}$, and
missing atoms $N_{\rm missing}$ for an isotropic trap (a) and 
an anisotropic trap (b).
[Reproduced from Fig. 4 of H. Saito and M. Ueda, Phys. Rev. A {\bf 65}, 033624
(2002).]
}
\label{f4}
\end{figure}

\bigskip
For an isotropic trap, the number of remnant BEC atoms is always 
found to be smaller than the critical number 
$N_{\rm cr}\simeq0.57 d_0/|a|$~\cite{Ruprecht}, as shown in 
Fig.~\ref{f4}(a).
However, for an anisotropic trap this is not necessarily the case,
because the burst atom cloud cannot refocus at the center of the BEC as 
efficiently as for an isotropic trap. 
For the cigar-shaped geometry used in Ref.~\cite{Donley}, we find that 
the fraction of the remnant BEC atoms does not show a significant 
dependence on the initial number of BEC atoms $N_0$ 
as shown in Fig.~\ref{f4}(b), in agreement with the observation in 
Ref.~\cite{Donley}.

It should be noted that the time scale of the intermittent implosions is
determined by the competition between the loss and self-focusing of 
the atoms, and is much shorter than the time scale of the inverse trap 
frequency $\omega^{-1}$.
The oscillatory behavior in the peak density associated with intermittent implosions should be distinguished from other oscillatory behaviors, such as the
collapse-and-growth cycle~\cite{Sackett,Kagan98} (which is the oscillation in
the number of BEC atoms) and small collapses found in Ref.~\cite{Kagan98}, 
both of which time scales are much longer than that of the intermittent 
implosions.

\begin{figure}[t]
\begin{center}
\includegraphics[width=8cm]{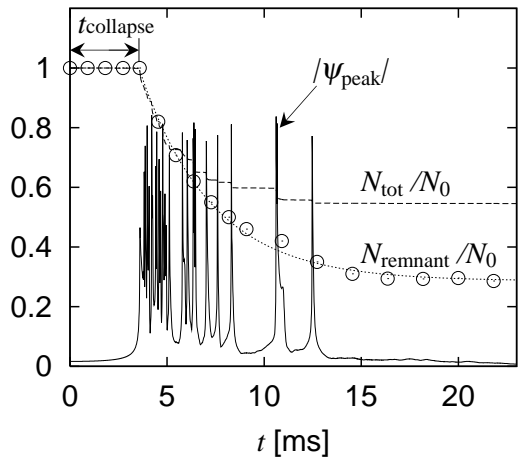}
\end{center}
\caption{
Dependences on time of the number of remnant BEC atoms
(dotted curve) and of the number of the remnant plus burst atoms 
(dashed curve).
[Reproduced from Fig. 1 of H. Saito and M. Ueda, Phys. Rev. A {\bf
65}, 033624 (2002).]
}
\label{f5}
\end{figure}

Perhaps the most convincing evidence of the intermittent implosions
is the gradual decay in the number of remnant BEC atoms 
(Fig.~1 of Ref.\cite{Donley}).
The dotted curve in Fig.~\ref{f5} shows the decay in the number of 
remnant BEC atoms, which is in excellent agreement with the experimental
observation~\cite{Donley}.
Our simulations thus reveal that the gradual decay is caused
by a series of intermittent implosions.

\begin{figure}[t]
\begin{center}
\includegraphics[width=8cm]{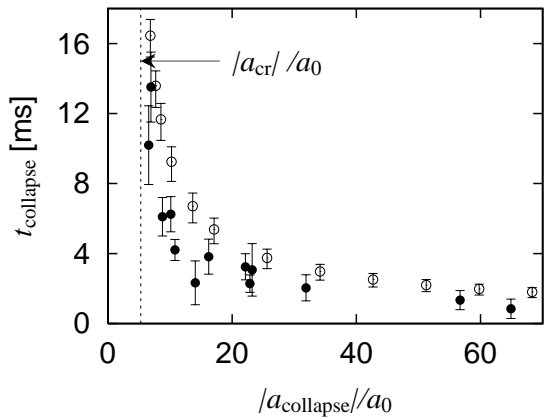}
\end{center}
\caption{
Dependence of $t_{\rm collapse}$ on $a_{\rm collapse}$.
Open circles show the results of our simulations, where the ^^ ^^ error bars" 
allow for experimental uncertainties of the initial value of the s-wave scattering length, $a_{\rm init}$~\cite{Donley}. Filled circles show the experimental results of the JILA group~\cite{Claussen_PhD}.
}
\label{f6}
\end{figure}

As shown in Fig.~\ref{f5}, $N_{\rm remnant}$ remains constant for some time, which is denoted as $t_{\rm collapse}$, after the switch of interactions. 
Open circles in Fig.~\ref{f6} show $t_{\rm collapse}$ obtained by our numerical simulations as a function of $a_{\rm collapse}$~\cite{SaitoPRA02}, 
where ^^ ^^ error bars" allow for experimental uncertainties $\pm 2a_0$, where $a_0$ is the Bohr radius, of the initial value $a_{\rm init}=0$ of the s-wave scattering length (see Ref.~\cite{Donley} for details); 
the upper and lower end points of each error bar correspond to $a_{\rm init}=-2a_0$ and $a_{\rm init}=2a_0$, respectively.
Filled circles show the latest experimental results of the JILA group~\cite{Claussen_PhD}. The agreement looks excellent.

\section{Atom Jets}

In the JILA experiments, prolonged atomic clouds, 
referred to in Ref.~\cite{Donley} as
^^ ^^ jets", were observed when the collapse was interrupted by switching 
the sign  
of interaction from attractive to repulsive.
The jets are distinguished from the burst in that the directions of 
the jets are purely radial.

\begin{figure}[t]
\begin{center}
\includegraphics[width=8cm]{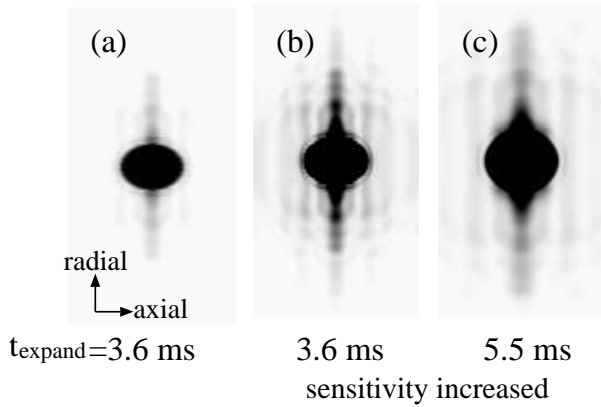}
\end{center}
\caption{
The integrated column densities seen from the radial direction
under the same conditions as the JILA experiments~\cite{Donley}.
[Reproduced from Fig. 6 of H. Saito and M. Ueda, Phys. Rev. A {\bf
65}, 033624 (2002).]
}
\label{f7}
\end{figure}

To understand this phenomenon, we performed numerical simulations 
under conditions similar to those of the experiment. 
Figure~\ref{f7} shows the gray-scale images of the integrated column
density after a $t_{\rm expand}=3.6$ ms expansion in (a) and 5.5 ms   
expansion in (c). These results look very similar to the experimental  
ones~\cite{Donley}.

\begin{figure}[t]
\begin{center}
\includegraphics[width=8cm]{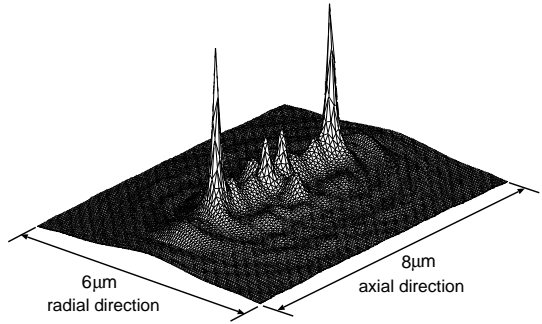}
\end{center}
\caption{
Snapshot of the collapsing BEC just before the collapse is
interrupted.
[Reproduced from Fig. 2 of H. Saito and M. Ueda, Phys. Rev. A {\bf
65}, 033624 (2002).]
}
\label{f8}
\end{figure}

The parallel fringe pattern seen in Fig.~\ref{f7}
is reminiscent of the interference fringe 
between waves emanating from two point sources. 
In fact, at the time when the interaction is switched from attractive to
repulsive, the collapsing condensate exhibits two spikes 
in the atomic density 
that serve as two point sources of matter waves,
 as shown in Fig.~\ref{f8}.

\medskip
{\it Thus the experimental observations of jets strongly suggests that 
atoms emanating from the spikes are coherent.}

\section{Pattern Formation}

We next consider a situation in which a BEC initially having a repulsive 
interaction is prepared and then the sign of interaction is 
suddenly switched to an attractive one. 

\begin{figure}[t]
\begin{center}
\includegraphics[width=8cm]{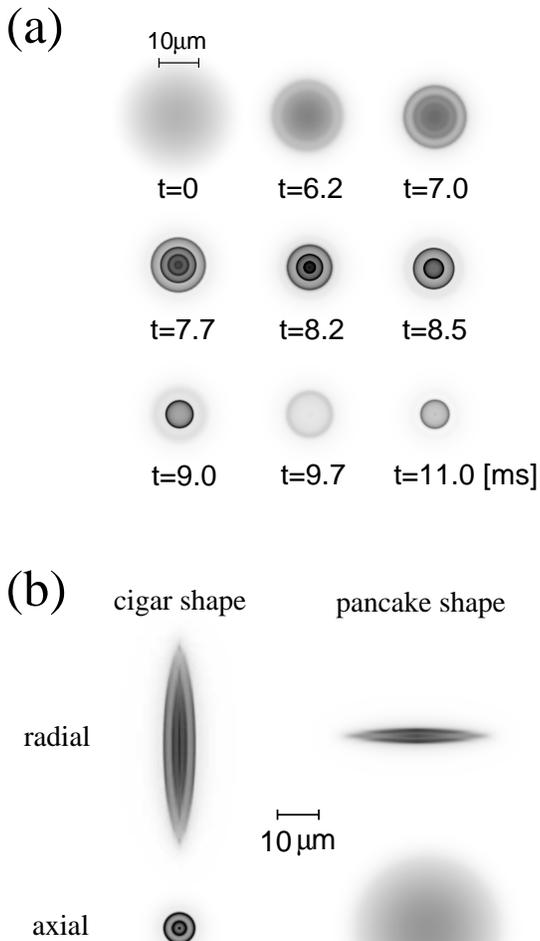}
\end{center}
\caption{
Time evolutions of the integrated column densities of three 
collapsing BECs: 
(a) isotropic trap, (b) cigar-shaped trap, and (c) pancake-shaped trap.
[Reproduced from Figs. 7 and 8 of H. Saito and M. Ueda, Phys. Rev. A {\bf
65}, 033624 (2002).]
}
\label{f9}
\end{figure}

Figure~\ref{f9} shows the time evolution of the integrated column density
of the collapsing BEC~\cite{SaitoPRL01,SaitoPRA02}.
We initially prepare a repulsive BEC with fifty thousands atoms, and at time
$t=0$ switch the sign of the interaction from repulsive to attractive.

We can see that as time elapses, shells of high atomic density
gradually form.
At time $t=7.7$ms, four shells are formed. 
But shortly after this formation ($t=8.2$ms),
the innermost shell collapses and disappears.
Then, one by one, the other shells collapse and disappear as they reach 
the center of the trap, until finally all the shells have disappeared 
completely.
The concentric circles in this gray-scale image clearly show the 
formation of the shell structure of the atomic density.
This reveals a new self-focusing effect of the attractive BEC.

For the cigar-shaped trap a cylindrical-shell structure is 
formed [Fig.~\ref{f9}(b)], while for the pancake-shaped trap 
a layered structure is formed [Fig.~\ref{f9}(c)].

\section{Collapse of attractive BECs with a vortex}

Finally, we discuss the collapse of attractive BECs with a vortex
\cite{SaitoPRL02}. 
This subject is of interest because, generally, 
attractive interactions cause the atoms to gather at the center, 
but in this case such gathering will be prohibited because the density 
right at the center must be    
zero due to the topological constraint of the vortex.
We thus expect an interesting competition between the self-focusing effect 
and the topological constraint.

Unfortunately, attractive BECs cannot hold vortices in any 
thermodynamically stable state.
However, the Feshbach technique enables us to prepare a vortex in an 
attractive BEC by first creating a vortex in a repulsive BEC and then    
switching the sign of interaction to attractive.
What are the ensuing dynamics of this system?

To study this problem we work with the GP 
equation~(\ref{GP}) in two dimensions,
and assume that the initial state is a single vortex state.
When the strength of attractive interactions exceeds a critical value,
the Bogoliubov spectrum for the quadrupole mode acquires an imaginary part. 
This signals the onset of dynamical instability of the quadrupole mode.
By increasing the strength of attractive interaction, we obtain a series of 
dynamical instabilities consisting of 
quadrupole, dipole, and monopole instabilities.

\begin{figure}[t]
\begin{center}
\includegraphics[width=8cm]{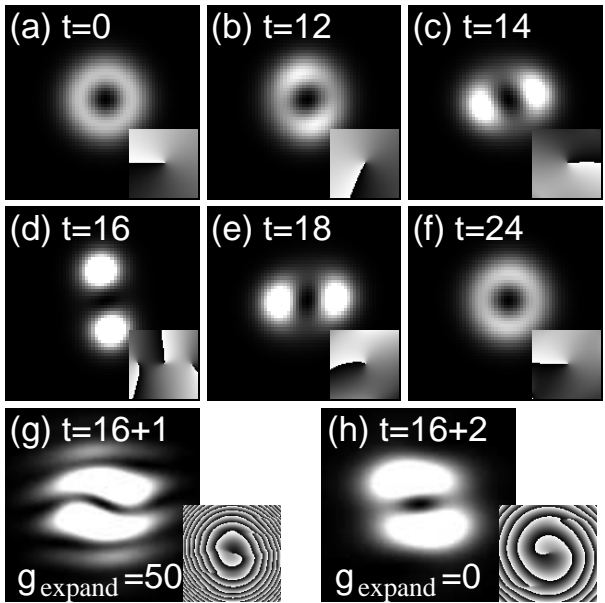}
\end{center}
\caption{
Split-merge cycle of a vortex with attractive interaction (a)-(f) and expanded
images (g) and (h).
[Reproduced from Fig. 3 of H. Saito and M. Ueda,
Phys. Rev. Lett. {\bf 89}, 190402 (2002).]
}
\label{f10}
\end{figure}

Figures~\ref{f10}(a)-(f) show the time evolution of the density and phase profiles
of a vortex state with  $g=-9$, which exceeds the critical value for the
quadrupole mode, but does not exceed that for the dipole mode.
Due to the quadrupole instability, the vortex splits into two clusters 
that rotate around the center of the trap, the two clusters
reunite to recover the original shape, and this split-merge 
cycle repeats.

The insets show the phase plots. At $t=16$, three branch cuts appear.
The central one exists from the outset, and the other two emerge as 
the vortex splits, in accordance with the fact that the $m=3$ component 
grows upon the vortex split. 
The two side vortices cannot be seen in the density plot, and thus
they may be considered ^^ ^^ ghost vorticies" that carry a 
negligible fraction of the angular momentum.

\begin{figure}[t]
\begin{center}
\includegraphics[width=8cm]{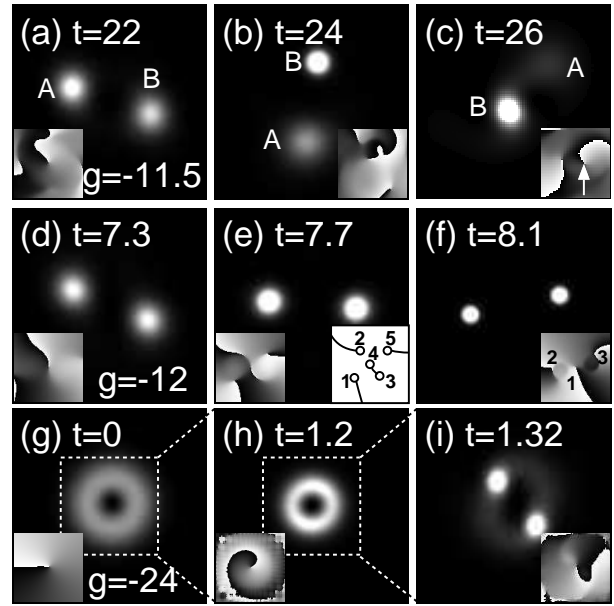}
\end{center}
\caption{
Collapse of a vortex when the dominant unstable mode is 
the dipole mode (a)-(f)
and the monopole mode (g)-(i).
[Reproduced from Fig. 4 of H. Saito and M. Ueda,
Phys. Rev. Lett. {\bf 89}, 190402 (2002).]
}
\label{f11}
\end{figure}

When the size of the system becomes too small to be observed by the 
{\it in situ} imaging, we may expand the condensate by switching the   
sign of interaction to positive. 
Figures~\ref{f11}(g) and (h) show snapshots of 
such expanded images. 
We find that the stronger the repulsive interactions, the more fringes 
appear.

Figure~\ref{f11} shows the case in which $g$ exceeds the critical value
for the dipole instability. The dipole instability gives rise to a
transferring of the atoms from one cluster to the other, thereby
inducing the collapse of the condensate.
In Figs.~(a) to (c), cluster A grows, and then cluster B grows 
in a seesawing motion, until finally cluster B 
has absorbed most of the atoms and collapsed.
In the case of (d)-(f), both clusters collapse immediately after the  
vortex split. In this collapse process, we find the exchange of a       
vortex-antivortex pair. 
Figures (g) to (i) show the case with even stronger attractive interactions,
in which the monopole instability appears.
We see that the vortex first shrinks isotropically, then
splits into two pieces, and finally collapses.

\section{Summary and Remarks}

We conclude this paper by summarizing our answers to each of the questions
presented in the abstract.

\medskip
\begin{itemize}

\item
BEC can survive partially after the collapse because the collapse occurs only in extremely localized regions and nowhere else in the BEC.

\item
For an isotropic trap, the number of remnant BEC atoms is always below the critical number.
However, for an anisotropic trap the number exceeds the corresponding 
critical number, since after the intermittent implosions the shape of the gas 
becomes highly distorted and the collapse does not necessarily occur at the center of a BEC. 

\item
The atom burst is caused by a slight surplus of the zero-point kinetic energy over the attractive interaction due to the removal of atoms in the highly localized high-density region by the three-body recombination loss.
Note that in the course of the collapse, the attractive interaction surpasses the kinetic energy; the sudden removal of atoms due to the loss breaks the balance slightly in favor of the kinetic energy, leading to a mean burst energy of as low as 100nK.

\item
The gradual decay in the number of remnant BEC atoms, reported in Fig.~1 of 
Ref.~\cite{Donley}, strongly suggests that the intermittent implosions
occur during the decay process. 

\item
A sudden interruption of the collapse by switching the sign of the interaction from attractive to repulsive leaves the collapsing BEC highly localized at two points for the cigar-shaped geometry under the conditions used in Ref.~\cite{Donley}. 
These two localized high-density regions subsequently play the role of point sources of matter waves
and yield the observed parallel fringe patterns. 
This explains why the jets occur only in the radial directions, unlike the bursts, and suggests that the jets are coherent.

\end{itemize}

\bigskip
It is interesting to observe that the atom burst also occurs when the
positive s-wave scattering length is substantially increased in a very
short time~\cite{Claussen,Donley02}.
The difference is that in these experiments the magnetic field is ramped 
up very close to the Feshbach resonance, where bound-state molecules are
likely to be formed. In the Bosenova experiments, in contrast, the magnetic
field is ramped away from the resonance, and therefore molecular formation
is expected to play a much smaller role than in the case of repulsive 
interaction. 
However, it seems likely that the two-body correlation is significantly
enhanced in the course of collapse. 
Moreover, in the midst of the collapse the peak density becomes so high 
that the quantum depletion of the condensate probably cannot be ignored.
Nonetheless, our extensive study described in this paper suggests that  
a generalized GP equation~(\ref{GP_loss}) together with the 
formula (\ref{mbe}) offers a consistent overall picture of the 
Bosenova experiments.

\bigskip
We acknowledge E. A. Donley and C. Wieman for valuable comments.
This work was supported by a Grant-in-Aid for Scientific Research
(Grant No.15340129) and Special Coordination Funds for Promoting 
Science and Technology by the Ministry of Education, Science, Sports,
and Culture of Japan, and by the Toray Science Foundation, and 
by the Yamada Science Foundation.

\end{document}